\documentclass[journal, onecolumn, 11pt]{IEEEtran}
\IEEEoverridecommandlockouts
\usepackage{cite}
\usepackage{amsmath,amssymb,amsfonts}
\usepackage{tikz}
\usetikzlibrary{positioning, arrows.meta, fit, backgrounds, calc}
\usepackage{algorithmic}
\usepackage{graphicx}
\usepackage{textcomp}
\usepackage{xcolor}
\usepackage{booktabs}
\usepackage[margin=1in, top=1.5in, bottom=1.5in]{geometry}
\usepackage{makecell}
\usepackage{lmodern}
\usepackage{placeins}

\usepackage[hidelinks]{hyperref}
\def\BibTeX{{\rm B\kern-.05em{\sc i\kern-.025em b}\kern-.08em
    T\kern-.1667em\lower.7ex\hbox{E}\kern-.125emX}}
    
\renewcommand{\thesection}{\arabic{section}}
\makeatletter
\renewcommand{\@seccntformat}[1]{\csname the#1\endcsname\quad}
\makeatother
\AtBeginDocument{%
  \renewcommand{\thesubsection}{\thesection.\arabic{subsection}}%
  \renewcommand{\thetable}{\arabic{table}}%
}

\begin{document}

\title{Citation-Driven Multi-View Training for Patent Embeddings: QaECTER and Sophia-Bench \\
}

\author{
  Younes Djemmal\textsuperscript{1},\quad
  You Zuo\textsuperscript{1,3},\quad
  Kim Gerdes\textsuperscript{2,4},\quad
  Kirian Guiller\textsuperscript{1}
  \thanks{\textsuperscript{1}AI Lab, Questel, Paris, France}
  \thanks{\textsuperscript{2}Qatent, Paris, France}
  \thanks{\textsuperscript{3}Inria, Paris, France}
  \thanks{\textsuperscript{4}Université Paris-Saclay, Orsay, France}
  \thanks{Email addresses: ydjemmal@questel.com, yzuo@questel.com, gerdes@lisn.fr, kguiller@questel.com}
}

\maketitle
\thispagestyle{empty}
\renewenvironment{abstract}{%
  \centering\small
  \textbf{Abstract}\par\medskip
  \begin{minipage}{0.9\linewidth}
}{%
  \end{minipage}\par\bigskip
}
\begin{abstract}
Patent retrieval underpins critical decisions in innovation, examination, and IP strategy, yet progress has been hampered by the absence of benchmarks that reflect the diversity of real-world search scenarios. We address this gap with two contributions. First, we introduce \textsc{Sophia-bench}, a large-scale patent retrieval benchmark comprising 10,000 queries and 75,000 corpus documents stratified across ten years, eight IPC technology sections, and twelve filing jurisdictions. Unlike prior benchmarks, \textsc{Sophia-bench} tests retrieval using 12 different query types-from structured patent fields to AI-generated summaries-and evaluates results against citation-based ground truth enhanced with a novel domain-relevance metric (InScope). Together, these enable systematic measurement of how well models perform across query types, technology domains, and jurisdictions. Second, we introduce \textsc{QaECTER}, a 344M-parameter embedding model trained on patent citation graphs and multi-view self-alignment. Despite its compact size, QaECTER establishes a new state of the art for patent retrieval. It outperforms the \#1 model on the English retrieval text embedding benchmark (RTEB)~\cite{RTEB}, a model 23$\times$ larger, as well as all existing patent-specific models across every query type, IPC section, and jurisdiction on Sophia-bench, with gains of up to 7.2\% average NDCG@10 over the next-best model. These results are confirmed on an independent external benchmark, where \textsc{QaECTER} surpasses all prior models without requiring task-specific instruction prompts. Both the benchmark and the model are designed for practical deployment in large-scale patent search systems.

\end{abstract}
\newpage
\section{Introduction}
\label{sec:introduction}

Patent search is central to the work of inventors, examiners, and IP attorneys. Before filing an application, an inventor must determine whether the idea is novel. An examiner receiving that application must locate relevant prior art to make a proper determination. An attorney advising on freedom-to-operate needs to identify patents that might block a product launch. In all cases, the quality of the search results matters: missing a critical reference can lead to wasted R\&D, invalidated patents, or costly litigation.

Patent retrieval presents fundamental challenges along four dimensions:

\begin{itemize}
    \item \textbf{Query diversity.} Search scenarios vary substantially depending on the practitioner's role and task. An inventor submitting a full invention disclosure, a patent examiner working from the claims of a pending application, and a landscape analyst starting from a problem statement or keyword set each impose distinct retrieval demands. The increasing use of AI-generated technical summaries further broadens this range. A system that performs well on one query type may degrade significantly on others, yet operational settings require robust performance across all of them.

    \item \textbf{Document complexity.} Patent documents constitute a hybrid of technical disclosure and legal instrument, with heterogeneous sections that serve distinct communicative functions. Claims are drafted in deliberately broad language to maximize legal scope; descriptions interleave technical content with extensive background and enumerated embodiments. The pronounced linguistic divergence across these sections means that effective retrieval cannot rely on surface-level matching but must capture how each section relates to the underlying invention.

    \item \textbf{Jurisdictional variation.} Structural and linguistic properties of patent documents vary systematically across jurisdictions, reflecting differences in claim dependency conventions, fee structures, and examination procedures. Retrieval systems must account for these cross-jurisdictional patterns to maintain consistent performance across heterogeneous patent offices.

    \item \textbf{Temporal coverage.} The global patent corpus spans several decades and grows by millions of applications annually. Technical vocabulary shifts considerably over time, and entire domains—transformer architectures, CRISPR-based gene editing, and autonomous vehicles—were nascent or nonexistent two decades ago. A practical retrieval system must bridge this temporal gap, surfacing relevant prior art using contemporary terminology while adapting to fields whose technical language remains in flux.
\end{itemize}

Given these challenges, one would expect a well-developed ecosystem of retrieval
benchmarks. In practice, existing resources either target patent-adjacent tasks
rather than retrieval, or impose constraints that limit their applicability to
real-world search scenarios:

\begin{itemize}
    \item \textbf{CLEF-IP}~\cite{clefip}: provides a large European patent benchmark for prior-art candidate search, but it is based on an old corpus and limited to three languages, reducing its representativeness and suitability for broader cross-lingual or global evaluation.

    \item \textbf{TREC-Chem}~\cite{trecchem}: restricted to chemical patents and
    journal articles, limiting both domain and document type coverage.

    \item \textbf{NTCIR}~\cite{ntcir}: earlier editions (NTCIR-4–6) supported patent invalidity search over collections from the Japan Patent Office and, in one English subtask, the United States Patent and Trademark Office, but later editions shifted away from retrieval tasks. Consequently, invalidity search is no longer maintained and remains limited to a small number of jurisdictions, reducing its suitability for general prior-art retrieval evaluation.

    \item \textbf{PatentMatch}~\cite{patentmatch}: frames prior art search as
    binary claim--passage classification rather than full-corpus retrieval.

    \item \textbf{DAPFAM}~\cite{dapfam}: the closest work to ours, introducing
    cross-domain retrieval with section-level queries. However, the query set is
    small (1{,}247 families) and concentrated in 2000--2015, limiting coverage of
    recent technological developments.

    \item \textbf{MTEB}~\cite{RTEB}: the dominant general-purpose embedding
    benchmark; includes legal retrieval tasks but no patent retrieval task.
\end{itemize}

No existing benchmark combines diverse query types, broad temporal and
jurisdictional coverage, and domain-aware relevance labels in a single evaluation
framework for patent retrieval.

We make two contributions. First, we introduce \textbf{\textsc{Sophia-bench}}, a large-scale patent retrieval benchmark covering the range of query types practitioners actually use: patent sections (titles, abstracts, claims, descriptions), structured extractions (object of the invention, independent claims set, advantages and disadvantages), and AI-generated summaries. The benchmark is built on citation-based ground truth spanning multiple years, technology domains, and jurisdictions, and introduces multiple testing axes (Figure~\ref{fig:evaluation-tasks}) that enable measurement of both average performance and robustness across query types, IPC sections, and filing jurisdictions. Second, we introduce \textbf{\textsc\textsc{QaECTER}}, a 344M-parameter embedding model built specifically for patent retrieval. Despite its compact size, it outperforms state-of-the-art general-purpose embedding models up to 23$\times$ larger, as well as existing patent-specific models, across query types, domains, and time periods on both \textsc{Sophia-bench} and external benchmarks, establishing a new state of the art for patent search.
% ══════════════════════════════════════════════════════════
%  SECTION 2 — qaecter
% ══════════════════════════════════════════════════════════

\section{Introducing \textsc{Qaecter}}
\label{sec:qaecter}

\subsection{Overview}
\label{sec:training-paradigm}

QaECTER is a 344M-parameter embedding model trained specifically for patent retrieval. The core training paradigm jointly optimizes two objectives: prior-art search and alignment of different \emph{patent views}—distinct textual representations of the same patent, such as the abstract, claims, description, or structured extractions like the object of the invention.

The model is trained on a large-scale dataset drawn from Questel's Orbit database\footnote{Questel, \textit{Orbit Intelligence -- Patent Search \& Analytics Software}, 2026. \url{https://www.questel.com/patent/ip-intelligence-software/orbit-intelligence/}}, one of the largest commercially available patent databases in the world. Training data spans patent applications published from 2002 to 2024, aggregated at the patent family level, and includes both original English text and translated English versions, making the model robust to linguistic and translation variation across jurisdictions.

Training pairs are derived from examiner citation relationships (X, Y and A categories), which provide complementary signals of semantic relevence at different levels of specificity. In addition, a proportion of training data consist of \emph{self-alignment pairs}: two different views of the same patent, ensuring that different textual facets of the same invention map to nearby points in the embedding space. This multi-view alignment is a key design choice, continuously exposing the model to heterogeneous representations of the same underlying inventive concept.

The evaluation benchmark (\textsc{Sophia-bench}, Section~\ref{sec:Sophia-bench}) was drawn from a 10-year window (2016--2025) with no overlap with the training set.

\subsection{Architecture}
\label{sec:architecture}

QaECTER builds on bert-for-patents~\cite{bertforpatents}, a BERT-Large model pre-trained via masked language modeling on a large-scale patent corpus. The tokenizer vocabulary is extended with special section tokens that delimit patent views during training, helping the model distinguish between section types; these tokens are not required at inference.

A projection head is used during training and discarded at inference: the final model outputs 1024-dimensional $L_2$-normalized embeddings via mean pooling over token representations.

We adopt a contrastive learning framework based on the InfoNCE objective with in-batch negatives and large effective batch sizes. Training proceeds for a single epoch with early stopping based on retrieval metrics.

% ══════════════════════════════════════════════════════════
%  SECTION 3 — Sophia-bench
% ══════════════════════════════════════════════════════════

\section{Sophia-bench}
\label{sec:Sophia-bench}
% ── Sophia-bench Evaluation Framework figure ──────────────────────────
% Requires in preamble:
%   \usepackage{tikz}
%   \usetikzlibrary{arrows.meta}

\begin{figure*}[htbp]
\centering
\resizebox{0.9\textwidth}{!}{%
% ── color palette ────────────────────────────────────────────────────
\definecolor{pAmber}{RGB}{161,98,7}
\definecolor{subAmber}{RGB}{254,243,199}
\definecolor{borAmber}{RGB}{217,169,56}
\definecolor{dlistbg}{RGB}{255,251,235}
\definecolor{dlistborder}{RGB}{234,197,94}
\definecolor{dlisttext}{RGB}{120,53,15}
\definecolor{edgemain}{RGB}{148,163,184}
\definecolor{edgefine}{RGB}{186,198,210}
\definecolor{anncolor}{RGB}{148,163,184}
\begin{tikzpicture}[
x=1cm, y=1cm,
every node/.style={align=center, inner sep=4pt, font=\scriptsize},
% ── node styles ──────────────────────────────────────────────────
root/.style={
rounded corners=4pt, fill=pAmber,
text=white, font=\footnotesize\bfseries,
minimum height=9mm, minimum width=42mm,
},
subC/.style={
rounded corners=3pt, fill=subAmber,
draw=borAmber, line width=0.45pt,
font=\scriptsize\bfseries, text=pAmber,
minimum height=7mm, minimum width=24mm,
},
dlist/.style={
rounded corners=3pt, fill=dlistbg,
draw=dlistborder, line width=0.45pt,
inner sep=3.5pt,
font=\scriptsize\ttfamily, text=dlisttext,
},
ann/.style={font=\scriptsize\itshape, text=anncolor},
link/.style={
-{Stealth[length=2.2pt, width=2pt, round]},
edgemain, line width=0.7pt, rounded corners=2pt,
},
linkfine/.style={
-{Stealth[length=1.6pt, width=1.5pt, round]},
edgefine, line width=0.45pt,
},
]

% ═══════════════════════════════════════════════════════════════════
% Root
% ═══════════════════════════════════════════════════════════════════
\node[root] (root) at (0, 0) {10K Queries / 75K Corpus};

% ═══════════════════════════════════════════════════════════════════
% Level 1 — Three stratification dimensions
% ═══════════════════════════════════════════════════════════════════
\node[subC] (Dt) at (-5.5, -1.7) {Temporal Coverage};
\node[subC] (Dj) at ( 0.0, -1.7) {Jurisdictions Distribution};
\node[subC] (Di) at ( 5.5, -1.7) {Technology Coverage};
\draw[link] (root) -- (Dt);
\draw[link] (root) -- (Dj);
\draw[link] (root) -- (Di);

% ═══════════════════════════════════════════════════════════════════
% Level 2 — Details
% ═══════════════════════════════════════════════════════════════════
\node[dlist] (Dtl) at (-5.5, -3.2)
{2016\;--\;2025\\1,000 queries/year};
\node[dlist] (Djl) at ( 0.0, -3.2)
{CHI~~ENG\\JPN~~KOR\\GER~~Other};
\node[dlist] (Dil) at ( 5.5, -3.2)
{A~~B~~C~~D\\E~~F~~G~~H};
\draw[linkfine] (Dt) -- (Dtl);
\draw[linkfine] (Dj) -- (Djl);
\draw[linkfine] (Di) -- (Dil);

% ═══════════════════════════════════════════════════════════════════
% Annotations
% ═══════════════════════════════════════════════════════════════════
\node[ann] at (-5.5, -4.6) {10-year span};
\node[ann] at ( 0.0, -4.6) {6 major languages};
\node[ann] at ( 5.5, -4.6) {all 8 IPC sections};

\end{tikzpicture}%
}% end resizebox
\caption{Corpus design in \textsc{Sophia-bench}, comprising 10,000 patent queries and a 75,000-document corpus stratified across three dimensions: temporal coverage (2016--2025, with 1,000 queries per year), jurisdiction distribution (Chinese, English, Japanese, Korean, German, and other jurisdictions), and technology coverage (all eight IPC sections A--H).}
\label{fig:corpus-design}
\end{figure*}
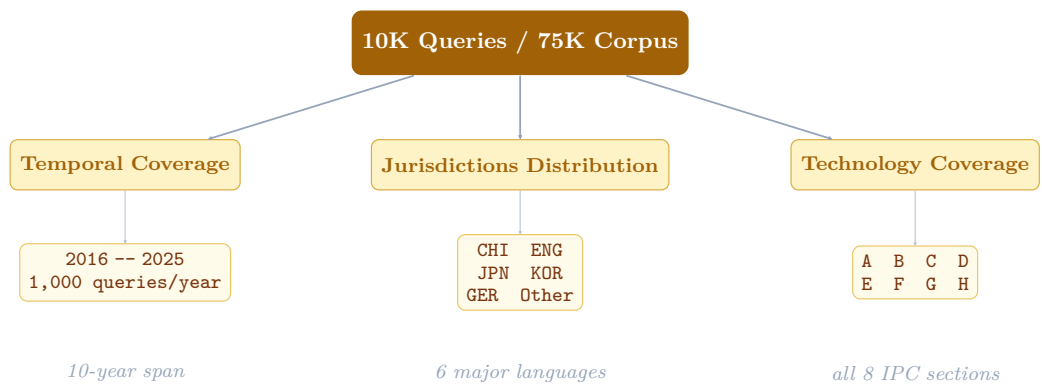

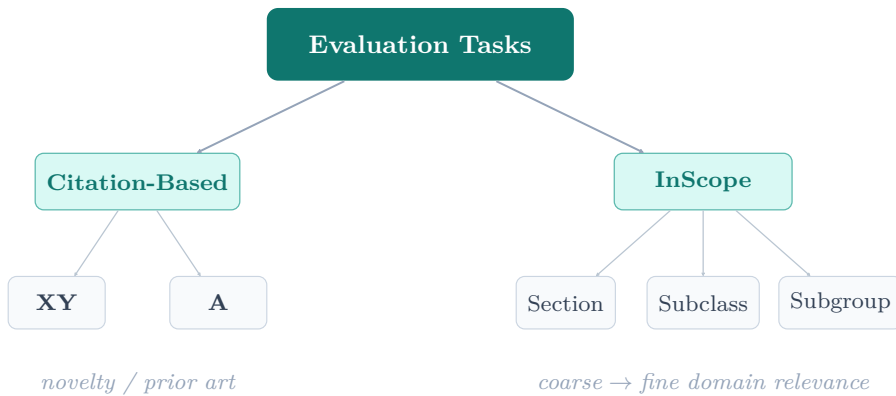
\begin{figure*}[htbp]
\centering
\resizebox{0.8\textwidth}{!}{%
% ── color palette ────────────────────────────────────────────────────
\definecolor{pTeal}{RGB}{15,118,110}
\definecolor{subTeal}{RGB}{217,249,244}
\definecolor{borTeal}{RGB}{94,186,175}
\definecolor{leafbg}{RGB}{248,250,252}
\definecolor{leafborder}{RGB}{203,213,225}
\definecolor{leaftext}{RGB}{51,65,85}
\definecolor{edgemain}{RGB}{148,163,184}
\definecolor{edgefine}{RGB}{186,198,210}
\definecolor{anncolor}{RGB}{148,163,184}
\begin{tikzpicture}[
x=1cm, y=1cm,
every node/.style={align=center, inner sep=4pt, font=\scriptsize},
% ── node styles ──────────────────────────────────────────────────
root/.style={
rounded corners=4pt, fill=pTeal,
text=white, font=\footnotesize\bfseries,
minimum height=9mm, minimum width=38mm,
},
subB/.style={
rounded corners=3pt, fill=subTeal,
draw=borTeal, line width=0.45pt,
font=\scriptsize\bfseries, text=pTeal,
minimum height=7mm, minimum width=22mm,
},
leaf/.style={
rounded corners=3pt, fill=leafbg,
draw=leafborder, line width=0.4pt,
font=\scriptsize, text=leaftext,
minimum height=6.5mm, minimum width=12mm,
},
ann/.style={font=\scriptsize\itshape, text=anncolor},
link/.style={
-{Stealth[length=2.2pt, width=2pt, round]},
edgemain, line width=0.7pt, rounded corners=2pt,
},
linkfine/.style={
-{Stealth[length=1.6pt, width=1.5pt, round]},
edgefine, line width=0.45pt,
},
]

% ═══════════════════════════════════════════════════════════════════
% Root
% ═══════════════════════════════════════════════════════════════════
\node[root] (root) at (0, 0) {Evaluation Tasks};

% ═══════════════════════════════════════════════════════════════════
% Level 1 — Two task types
% ═══════════════════════════════════════════════════════════════════
\node[subB] (Rc) at (-3.5, -1.7) {Citation-Based};
\node[subB] (Ri) at ( 3.5, -1.7) {InScope};
\draw[link] (root) -- (Rc);
\draw[link] (root) -- (Ri);

% ═══════════════════════════════════════════════════════════════════
% Level 2 — Task variants
% ═══════════════════════════════════════════════════════════════════
\node[leaf] (xy) at (-4.5, -3.2) {\textbf{XY}};
\node[leaf] (aa) at (-2.5, -3.2) {\textbf{A}};
\draw[linkfine] (Rc) -- (xy);
\draw[linkfine] (Rc) -- (aa);

\node[leaf] (s1) at ( 1.8, -3.2) {Section};
\node[leaf] (s4) at ( 3.5, -3.2) {Subclass};
\node[leaf] (sf) at ( 5.2, -3.2) {Subgroup};
\draw[linkfine] (Ri) -- (s1);
\draw[linkfine] (Ri) -- (s4);
\draw[linkfine] (Ri) -- (sf);

% ═══════════════════════════════════════════════════════════════════
% Annotations
% ═══════════════════════════════════════════════════════════════════
\node[ann] at (-3.5, -4.2) {novelty / prior art};
\node[ann] at ( 3.5, -4.2) {coarse\;$\rightarrow$\;fine domain relevance};

\end{tikzpicture}%
}% end resizebox
\caption{Evaluation tasks in \textsc{Sophia-bench}: citation-based retrieval evaluates novelty and prior-art search using XY (cited by examiner) and A (cited by applicant) citations, while InScope tasks measure domain relevance at three IPC granularities from coarse Section-level to fine Subgroup-level classification.}
\label{fig:evaluation-tasks}
\end{figure*}
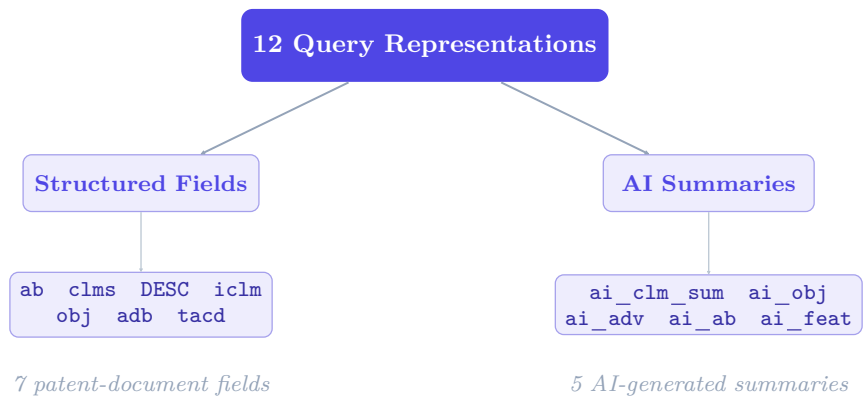
\begin{figure*}[t]
\centering
\resizebox{0.8\textwidth}{!}{%
% ── color palette ────────────────────────────────────────────────────
\definecolor{pIndigo}{RGB}{79,70,229}
\definecolor{subIndigo}{RGB}{238,237,253}
\definecolor{borIndigo}{RGB}{165,160,236}
\definecolor{leafbg}{RGB}{248,250,252}
\definecolor{leafborder}{RGB}{203,213,225}
\definecolor{leaftext}{RGB}{51,65,85}
\definecolor{qlistbg}{RGB}{238,237,253}
\definecolor{qlistborder}{RGB}{165,160,236}
\definecolor{qlisttext}{RGB}{55,48,163}
\definecolor{edgemain}{RGB}{148,163,184}
\definecolor{edgefine}{RGB}{186,198,210}
\definecolor{anncolor}{RGB}{148,163,184}
\begin{tikzpicture}[
x=1cm, y=1cm,
every node/.style={align=center, inner sep=4pt, font=\scriptsize},
% ── node styles ──────────────────────────────────────────────────
root/.style={
rounded corners=4pt, fill=pIndigo,
text=white, font=\footnotesize\bfseries,
minimum height=9mm, minimum width=38mm,
},
subA/.style={
rounded corners=3pt, fill=subIndigo,
draw=borIndigo, line width=0.45pt,
font=\scriptsize\bfseries, text=pIndigo,
minimum height=7mm, minimum width=26mm,
},
qlist/.style={
rounded corners=3pt, fill=qlistbg,
draw=qlistborder, line width=0.45pt,
inner sep=3.5pt,
font=\scriptsize\ttfamily, text=qlisttext,
},
ann/.style={font=\scriptsize\itshape, text=anncolor},
link/.style={
-{Stealth[length=2.2pt, width=2pt, round]},
edgemain, line width=0.7pt, rounded corners=2pt,
},
linkfine/.style={
-{Stealth[length=1.6pt, width=1.5pt, round]},
edgefine, line width=0.45pt,
},
]

% ═══════════════════════════════════════════════════════════════════
% Root
% ═══════════════════════════════════════════════════════════════════
\node[root] (root) at (0, 0) {12 Query Representations};

% ═══════════════════════════════════════════════════════════════════
% Level 1 — Two categories
% ═══════════════════════════════════════════════════════════════════
\node[subA] (Qs) at (-3.5, -1.7) {Structured Fields};
\node[subA] (Qa) at ( 3.5, -1.7) {AI Summaries};
\draw[link] (root) -- (Qs);
\draw[link] (root) -- (Qa);

% ═══════════════════════════════════════════════════════════════════
% Level 2 — Field lists
% ═══════════════════════════════════════════════════════════════════
\node[qlist] (Qsl) at (-3.5, -3.2)
{ab~~clms~~DESC~~iclm\\obj~~adb~~tacd};
\node[qlist] (Qal) at ( 3.5, -3.2)
{ai\_clm\_sum~~ai\_obj\\ai\_adv~~ai\_ab~~ai\_feat};
\draw[linkfine] (Qs) -- (Qsl);
\draw[linkfine] (Qa) -- (Qal);

% ═══════════════════════════════════════════════════════════════════
% Annotations
% ═══════════════════════════════════════════════════════════════════
\node[ann] at (-3.5, -4.2) {7 patent-document fields};
\node[ann] at ( 3.5, -4.2) {5 AI-generated summaries};

\end{tikzpicture}%
}% end resizebox
\caption{The twelve query representations in \textsc{Sophia-bench}, comprising seven structured patent-document fields (abstract, claims, description, etc.) and five AI-generated textual summaries (claim summary, objective, advantages, abstract summary, and features).}
\label{fig:query-representations}
\end{figure*}

\textsc{Sophia-bench} is a large-scale benchmark for evaluating patent embedding models on prior-art retrieval. Unlike existing benchmarks, \textsc{Sophia-bench} systematically varies the query text representation across 12~types while keeping the corpus fixed, enabling a controlled study of how different textual facets of a patent affect retrieval quality.

\subsection{Dataset Construction}
\label{sec:dataset-construction}

The benchmark is built from a corpus of 75,050 patent documents and a query set of 10,000 patents, both drawn from Questel's international patent database. All texts are in English: for patents originally filed in other languages, we use the English-language family representative (original or translated). Each corpus document is represented by the concatenation of its title, abstract, claims, and description (\texttt{tacd}).

The query set contains 1,000 patents per year from 2016 to 2025. The set spans all eight IPC sections (A--H), covering technology domains from physics and electricity to chemistry, mechanical engineering, and human necessities. In terms of original filing jurisdiction, the queries reflect the global distribution of patent activity across 12~jurisdictions, including Chinese, English-speaking, Japanese, Korean and European patent offices. This diversity in both domain and jurisdiction enables evaluation of model robustness across patents originating from different technological fields and jurisdictional conventions.

\subsection{Ground Truth}
\label{sec:ground-truth}

Relevance judgments are derived from patent citation graphs. For each query patent~$q$, the relevant set $\mathcal{R}_q$ is constructed from four citation relationships:
\begin{itemize}
    \item \textbf{Cited-XY}: patents cited by~$q$ as X or Y references (novelty or inventive step);
    \item \textbf{Citing-XY}: patents that cite~$q$ as X or Y references;
    \item \textbf{Cited-A}: patents cited by~$q$ as A references (general technological background);
    \item \textbf{Citing-A}: patents that cite~$q$ as A references.
\end{itemize}
The union of all four sets constitutes the full relevance set~$\mathcal{R}_q$. On average, each query has 6.6~relevant documents (median~5, range~2--90). All 10,000 queries have at least one XY-relevant document; 9,196 have at least one A-relevant document. These two citation categories represent different levels of semantic similarity and are evaluated separately (Section~\ref{sec:eval-dimensions}).

Each document is also annotated with its International Patent Classification (IPC) codes, enabling domain-level analysis of retrieval performance. The IPC taxonomy is hierarchical: section (one of eight broad domains, e.g., G = Physics), subclass (e.g., G06K = Recognition and presentation of data), and subgroup (e.g., G06Q-010/026).

\subsection{Query Representations}
\label{sec:query-representations}

A central axis of \textsc{Sophia-bench} is the systematic variation of query text representation. While the corpus encoding is fixed (\texttt{tacd}), we evaluate 12~query text types grouped into two families.

\paragraph{Structured patent fields.}
These correspond to standard patent sections or fields extracted via regular expressions:

\begin{itemize}
    \item \texttt{ab}: abstract;
    \item \texttt{clms}: full claims;
    \item \texttt{DESC}: full description;
    \item \texttt{iclm}: independent claims only;
    \item \texttt{obj}: extracted object of the invention;
    \item \texttt{adb}: extracted advantages and disadvantages;
    \item \texttt{tacd}: title + abstract + claims + description (same representation as the corpus).
\end{itemize}

\paragraph{AI-generated summaries.}
These are produced by a large language model applied to the full patent text, providing concise, semantically rich reformulations:
\begin{itemize}
    \item \texttt{ai\_claim\_summary}: generated claims summary;
    \item \texttt{ai\_obj}: generated object of the invention;
    \item \texttt{ai\_adv}: generated advantages;
    \item \texttt{ai\_ab}: generated abstract;
    \item \texttt{ai\_features}: generated technical features of the invention.
\end{itemize}

This design isolates the effect of query representation: by fixing the corpus encoding and varying only the query side, we can directly measure how much information each text type carries for retrieval, and how robust each model is across the query types.
\FloatBarrier

% ══════════════════════════════════════════════════════════
%  SECTION 4 — Evaluation Protocol
% ══════════════════════════════════════════════════════════

\section{Evaluation Protocol}
\label{sec:evaluation}

\subsection{Models}
\label{sec:models}

\begin{table}[t]
\centering
\caption{Embedding models evaluated on \textsc{Sophia-bench}. All models produce $L_2$-normalized embeddings and are evaluated with a maximum sequence length of 512~tokens.}
\label{tab:models}
\small
\begin{tabular}{lccr}
\toprule
\textbf{Model} & \textbf{Domain} & \textbf{Dim.} & \textbf{Params} \\
\midrule
Octen-Embed-8B       & General & 4096 &  8B  \\
Octen-Embed-4B       & General & 2560 &  4B  \\
PaECTER              & Patent  & 1024 & 344M \\
bert-for-patents     & Patent  & 1024 & 344M \\
patembed-large       & Patent  & 1024 & 344M \\
modernbert-embed-base & General &  768 & 100M \\
QaECTER (ours)   & Patent  & 1024 & 344M \\
\bottomrule
\end{tabular}
\end{table}

We evaluate seven embedding models spanning patent-specific and general-purpose architectures. Table~\ref{tab:models} summarizes their properties.

\textbf{Octen-Embed-4B} and \textbf{Octen-Embed-8B}~\cite{octen} are large-scale general-purpose embedding models based on Qwen3\cite{qwen3} decoder architectures, fine-tuned with LoRA for embedding tasks. \textbf{Octen-Embed-8B} ranks \#1 on the Hugging Face RTEB Leaderboard \cite{RTEB} and the 4B variant ranks \#2 (as of January 12, 2026).

\textbf{PAECTER}~\cite{paecter} is a patent-domain embedding model fine-tuned with a contrastive objective on patent citation pairs, built on BERT-Large for Patents.

\textbf{BERT-for-Patents}~\cite{bertforpatents} is a BERT-Large model pre-trained from scratch on patent text via masked language modeling; we wrap it with mean pooling for embedding extraction.

\textbf{PatEmbed-Large}~\cite{patembed} is an instruction-tuned patent embedding model built on BERT-for-Patents; following the authors' recommendations, we prepend the specified query and document prefixes for the task (retrieval mixed) for fair comparison.

\textbf{ModernBERT-Embed-Base}~\cite{modernbert} is a recent general-purpose embedding model based on a modernized BERT architecture.

\textbf{QaECTER} is our proposed model (Section~\ref{sec:qaecter}).

\subsection{Retrieval Setup}
\label{sec:retrieval-setup}

For each model, the entire corpus of 75,050 patent documents is embedded once using the \texttt{tacd} text representation, producing a single set of corpus vectors. Each query patent, however, can be represented using our 12~different text types, each yielding a distinct query embedding.

To measure how relevant each corpus document is to a given query, we use cosine similarity, a standard measure of how close two vectors are in the vector space. The score ranges from $-1$ (completely unrelated) to $1$ (a perfect match). The higher the score, the more semantically similar the two patents are according to the model.

For each query, all 75,050~corpus documents are scored and ranked in descending order of similarity, producing one complete ranking per query. Since we evaluate 7~models, each with 12~query text types, this yields $7 \times 12 = 84$~distinct retrieval configurations, allowing us to systematically compare how the choice of model and query representation affects retrieval quality.

All embeddings are computed with a maximum sequence length of 512~tokens. Texts exceeding this limit are truncated by the tokenizer. No special tokens or instruction prefixes are used except for the model PatEmbed-Large, which requires task-specific prefixes as specified by its authors.

\subsection{Metrics}
\label{sec:metrics}

We employ two families of metrics capturing complementary aspects of retrieval quality.

\subsubsection{Citation-Based Retrieval Metrics}
\label{sec:citation-metrics}
The objective here is to test how well the models can find the examiner's citations for each patent with the different types explained in (Section~\ref{sec:ground-truth}).

We use standard information retrieval metrics: Mean Reciprocal Rank (MRR), Mean Average Precision (MAP), Recall@$k$, and NDCG@$k$ computed against the citation-based ground truth at different cutoffs $k \in \{1, 5, 10, 20, 50, 100, 200, 500, 1\text{K}$).

\subsubsection{Domain Relevance: InScope}
\label{sec:inscope}

Citation-based metrics only credit retrieval of documents that appear in the citation graph. A model may retrieve semantically relevant patents that are not among the known citations available in the ground truth search reports. Such results are penalized by citation metrics but may still be relevant and useful to a practitioner. To capture this, we introduce the \textbf{InScope} metric, which measures the proportion of top-$k$ retrieved documents that share at least one IPC code with the query, truncated to a given granularity level~$g$:
\begin{equation}
    \text{InScope}_g\text{@}k = \frac{1}{|\mathcal{Q}|} \sum_{q \in \mathcal{Q}} \frac{|\{d \in D_k : \text{IPC}_g(d) \cap \text{IPC}_g(q) \neq \emptyset\}|}{k}
\end{equation}
where $D_k$ is the set of top-$k$ retrieved documents and $\text{IPC}_g(d)$ denotes the IPC codes of document~$d$ truncated to granularity~$g$.

In other words, InScope answers: \emph{among the top-ranked results, what fraction belongs to the same technological domain as the query patent?} We evaluate this at three levels of IPC granularity, from broad to narrow:
\begin{itemize}
    \item \textbf{Section}: one of eight broad domains (e.g., \texttt{G} = Physics);
    \item \textbf{Subclass}: a specific technology subclass (e.g., \texttt{G06Q});
    \item \textbf{Subgroup}: the most specific IPC level (e.g., \texttt{G06Q-010/026}).
\end{itemize}
InScope is computed at $k \in \{1, 5, 10, 20, 50, 100\}$. Higher values indicate that the model's top results are more topically concentrated in the correct domain.

\subsection{Evaluation Dimensions}
\label{sec:eval-dimensions}

All citation-based metrics are computed along four complementary axes:

\begin{enumerate}
    \item \textbf{Overall}: all queries, all relevant documents.
    \item \textbf{By citation category}: separate evaluation on XY~citations (novelty/inventive step) and A~citations (general background). These categories represent fundamentally different levels of semantic relatedness: XY references are closely related to the inventive concept, while A references provide broader technological context.
    \item \textbf{By IPC section}: per-domain evaluation across the eight IPC sections (A--H), revealing whether models exhibit domain-specific strengths or weaknesses.
    \item \textbf{By original jurisdiction}: per-jurisdiction evaluation across the 12~represented filing origins, assessing model robustness on patents originating from different patent office conventions and translation pipelines.
\end{enumerate}

% ══════════════════════════════════════════════════════════
%  SECTION 5 — Results
% ══════════════════════════════════════════════════════════

\section{Results and Discussion}
\label{sec:results}

QaECTER establishes a new state of the art for patent retrieval, surpassing all existing patent-specific and general-purpose embedding models known to us. We demonstrate this through extensive evaluation on \textsc{Sophia-bench}, covering 12~query types, 7~models, and multiple evaluation axes, as well as on an independent external benchmark. The following subsections present and discuss these results in detail.

% ──────────────────────────────────────────────────────────
\subsection{Citation-Based Retrieval}
\label{sec:citation-results}

We begin with citation-based retrieval, selecting the best-performing query type from each group of queries based on average NDCG@10: \texttt{tacd} for structured patent fields and \texttt{ai\_features} for AI-generated queries. Table~\ref{tab:citation-retrieval} reports Recall@10, Recall@100, and NDCG@10, broken down by citation category: XY (novelty/inventive step), A (general background), and all citations combined.

\begin{table*}[t]
\centering
\caption{Citation-based retrieval on \textsc{Sophia-bench}. Models are sorted by average NDCG@10. Best result per column is in \textbf{bold}.}
\label{tab:citation-retrieval}
\small
\setlength{\tabcolsep}{4pt}
\begin{tabular}{ll ccc ccc ccc}
\toprule
& & \multicolumn{3}{c}{\textbf{XY}} & \multicolumn{3}{c}{\textbf{A}} & \multicolumn{3}{c}{\textbf{All}} \\
\cmidrule(lr){3-5} \cmidrule(lr){6-8} \cmidrule(lr){9-11}
\textbf{Query} & \textbf{Model} & R@10 & R@100 & NDCG & R@10 & R@100 & NDCG & R@10 & R@100 & NDCG \\
\midrule
\texttt{tacd}
& QaECTER (ours)        & \textbf{.582} & \textbf{.873} & \textbf{.473} & \textbf{.486} & \textbf{.831} & \textbf{.371} & \textbf{.529} & \textbf{.849} & \textbf{.538} \\
& Octen-Embed-8B        & .536 & .836 & .436 & .452 & .798 & .350 & .491 & .813 & .502 \\
& Octen-Embed-4B        & .527 & .833 & .428 & .445 & .797 & .344 & .482 & .811 & .493 \\
& PaECTER               & .528 & .834 & .436 & .428 & .781 & .334 & .473 & .802 & .491 \\
& patembed-large        & .519 & .824 & .423 & .432 & .779 & .334 & .471 & .797 & .483 \\
& modernbert-embed-base & .383 & .684 & .315 & .324 & .639 & .258 & .352 & .659 & .368 \\
& bert-for-patents      & .279 & .521 & .234 & .208 & .452 & .172 & .240 & .481 & .261 \\
\midrule
\texttt{ai\_feat.}
& QaECTER (ours)        & \textbf{.577} & \textbf{.875} & \textbf{.472} & \textbf{.482} & \textbf{.833} & \textbf{.366} & \textbf{.527} & \textbf{.851} & \textbf{.535} \\
& Octen-Embed-8B        & .546 & .844 & .445 & .451 & .804 & .346 & .494 & .821 & .505 \\
& Octen-Embed-4B        & .532 & .838 & .434 & .441 & .798 & .338 & .482 & .814 & .493 \\
& PaECTER               & .511 & .825 & .424 & .417 & .772 & .322 & .461 & .795 & .477 \\
& patembed-large        & .507 & .822 & .411 & .424 & .780 & .323 & .461 & .797 & .469 \\
& modernbert-embed-base & .367 & .682 & .298 & .309 & .634 & .241 & .337 & .657 & .347 \\
& bert-for-patents      & .218 & .475 & .179 & .171 & .421 & .136 & .194 & .448 & .203 \\
\bottomrule
\end{tabular}
\end{table*}

QaECTER achieves the highest scores across all metrics, citation categories, and both query types. On the \texttt{tacd} query, it reaches an NDCG@10 of 0.538, outperforming the next-best model (Octen-Embed-8B, 0.502) by 3.6\%. This gap is notable given that Octen-Embed-8B has 8~billion parameters-23$\times$ more than QaECTER's 344~million. QaECTER outperforms existing patent embeddings models PaECTER and patembed-large by an even larger margin of 4.7\% and 5.5\% respectively.

Across all models, XY citations are consistently easier to retrieve than A citations, with higher recall and ranking scores. This is expected: XY references are closely tied to the query patent's core inventive concept, making them more semantically similar, whereas A references provide broader technological context, making them harder to distinguish from other patents in the same domain. QaECTER maintains the lead on both citation types across all general-purpose and patent embedding models.

The results are consistent across all query types. \texttt{tacd}, which uses the full patent text as query, achieves marginally higher scores than \texttt{ai\_features}, but the model ranking is identical. This confirms that the generalization of QaECTER to different query types, and more importantly, it showcases its superior search capabilities to other models when searching using the technical features of an invention. 

Figure~\ref{fig:recall-curves} shows the full Recall@$k$ curves for the \texttt{tacd} query type up to $k = 1{,}000$. QaECTER maintains a consistent lead at every cutoff, with the gap widening in the low-$k$ regime where precision matters most for practitioners. At $k = 10$, QaECTER retrieves 53\% of all relevant documents, compared to 49\% for Octen-Embed-8B and 47\% for PaECTER and patembed-large, a gap that is expected to grow larger when searching in larger production grade databases of patents.

\begin{figure*}[t]
\centering
\includegraphics[width=\textwidth]{fig_recall_curves.png}
\caption{Recall@$k$ curves on \textsc{Sophia-bench} for the \texttt{tacd} query type (overall). QaECTER (blue) leads at every cutoff from $k = 1$ to $k = 1{,}000$.}
\label{fig:recall-curves}
\end{figure*}

% ──────────────────────────────────────────────────────────
\subsection{Robustness Across Query Types}
\label{sec:query-robustness}

Table~\ref{tab:query-robustness} reports NDCG@10 for all 12~query types, providing a comprehensive view of how each model handles different textual representations of the same underlying inventions.

\begin{table}[t]
\centering
\caption{NDCG@10 across all 12 query types on \textsc{Sophia-bench} (overall). Best result per row in \textbf{bold}. Query types are grouped into structured patent fields (top) and AI-generated summaries (bottom).}
\label{tab:query-robustness}
\small
\setlength{\tabcolsep}{4pt}
\begin{tabular}{l @{\hspace{6pt}} c @{\hspace{4pt}} c @{\hspace{4pt}} c @{\hspace{4pt}} c @{\hspace{4pt}} c @{\hspace{4pt}} c @{\hspace{4pt}} c}
\toprule
\textbf{Query type} & \textbf{QaECTER} & \makecell{\textbf{Octen-}\\\textbf{Embed-8B}} & \makecell{\textbf{Octen-}\\\textbf{Embed-4B}} & \textbf{PaECTER} & \makecell{\textbf{patembed-}\\\textbf{large}} & \makecell{\textbf{modernbert-}\\\textbf{embed-base}} & \makecell{\textbf{bert-for-}\\\textbf{patents}} \\
\midrule
\texttt{tacd}         & \textbf{.538} & .502 & .493 & .491 & .483 & .368 & .261 \\
\texttt{clms}         & \textbf{.524} & .490 & .476 & .469 & .465 & .335 & .194 \\
\texttt{ab}           & \textbf{.517} & .480 & .471 & .474 & .459 & .358 & .234 \\
\texttt{iclm}         & \textbf{.517} & .482 & .471 & .455 & .461 & .344 & .207 \\
\texttt{obj}          & \textbf{.474} & .441 & .431 & .419 & .424 & .334 & .185 \\
\texttt{DESC}         & \textbf{.481} & .446 & .421 & .403 & .412 & .343 & .092 \\
\texttt{adb}          & \textbf{.383} & .348 & .330 & .322 & .333 & .247 & .103 \\
\midrule
\texttt{ai\_ab}       & \textbf{.539} & .502 & .493 & .480 & .472 & .393 & .139 \\
\texttt{ai\_feat}     & \textbf{.535} & .505 & .493 & .477 & .469 & .347 & .203 \\
\texttt{ai\_clm\_sum} & \textbf{.537} & .501 & .492 & .480 & .476 & .325 & .197 \\
\texttt{ai\_adv}      & \textbf{.496} & .465 & .453 & .425 & .433 & .317 & .078 \\
\texttt{ai\_obj}      & \textbf{.482} & .456 & .440 & .418 & .420 & .351 & .093 \\
\midrule
\emph{Average}        & \textbf{.502} & .468 & .455 & .443 & .442 & .339 & .166 \\
\emph{Max\,--\,Min}   & .156 & .157 & .163 & .169 & .150 & .146 & .183 \\
\bottomrule
\end{tabular}
\end{table}

QaECTER ranks first on all 12~query types, with an average NDCG@10 of 0.502. The margin over the second-best model (Octen-Embed-8B, 0.468) is consistent across query families, indicating a systematic advantage rather than one driven by specific text types. We can particularly see the superiority of QaECTER over the patent specific embeddings models here by looking at the low generalization capabilities of PaECTER and patembed-large that suffer from a high variability in performance across query types. 

The performance spread across query types (max minus min NDCG@10) is similar for the top models (0.15--0.17), with \texttt{adb} (advantages and disadvantages) being the most challenging and \texttt{ai\_ab} or \texttt{tacd} the easiest. The difficulty of \texttt{adb} likely stems from its shorter length and narrower semantic scope compared to full-text representations.

AI-generated summaries perform comparably to their structured counterparts. For instance, \texttt{ai\_ab} (0.539) slightly outperforms \texttt{ab} (0.517) for QaECTER, suggesting that the LLM reformulation adds useful semantic signal. Similarly, \texttt{ai\_claim\_summary} (0.537) outperforms \texttt{clms} (0.524). This finding validates the use of AI-generated queries in patent search workflows.

bert-for-patents exhibits the largest performance spread (0.183) and near-zero scores on \texttt{DESC} and \texttt{ai\_adv}, confirming that without task-specific fine-tuning, general patent language modeling does not transfer effectively to retrieval.

% ──────────────────────────────────────────────────────────
\subsection{Jurisdiction and Temporal generalisation}
\label{sec:jurisdiction-results}

\begin{figure*}[t]
\centering
\includegraphics[width=\textwidth]{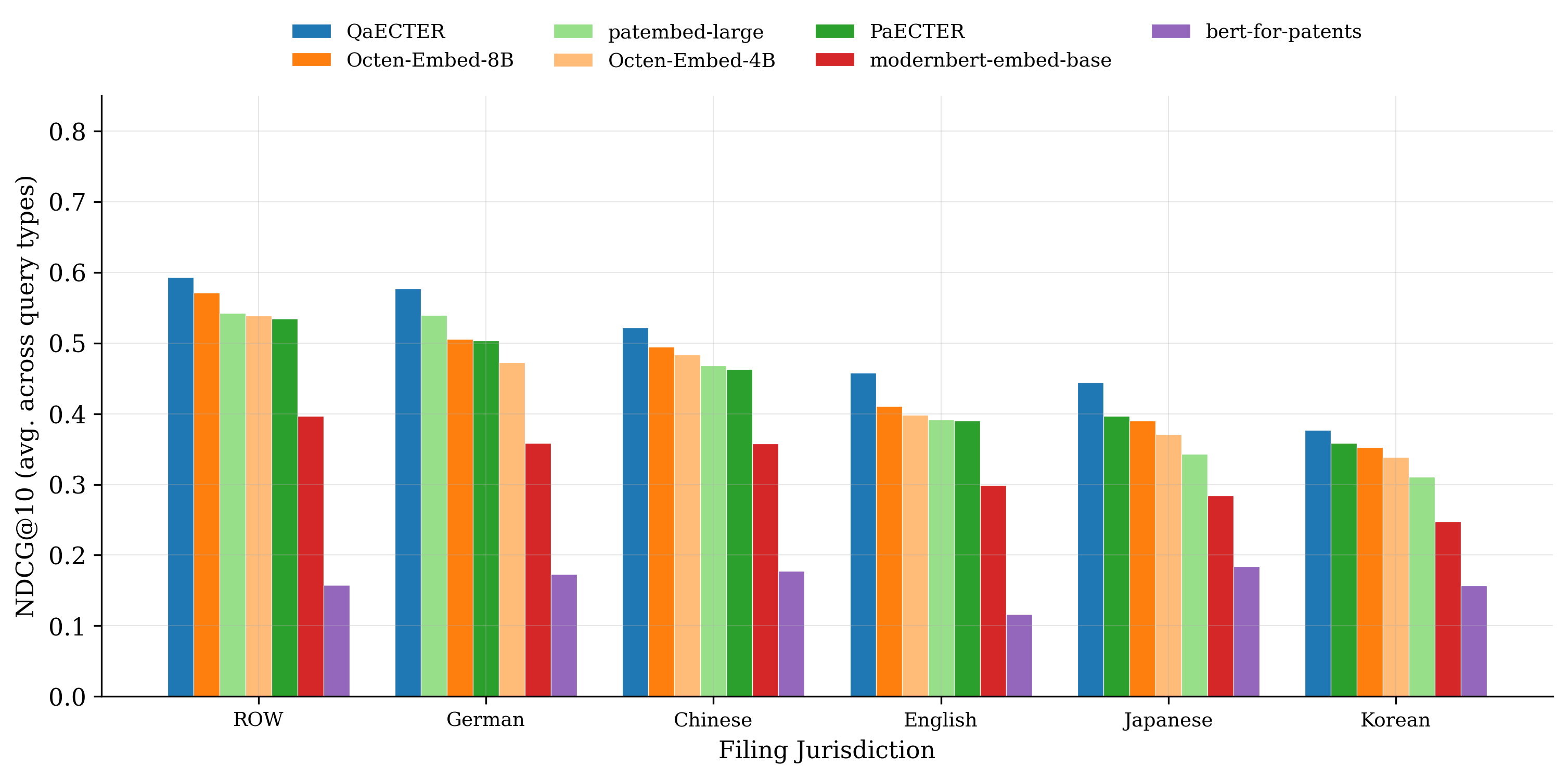}
\caption{NDCG@10 by filing jurisdiction on \textsc{Sophia-bench}, averaged across all 12 query types. Jurisdictions are sorted by QaECTER performance (descending). Four jurisdictions with fewer than five queries are excluded.}
\label{fig:ndcg-by-apl}
\end{figure*}

Figure~\ref{fig:ndcg-by-apl} shows NDCG@10 averaged across all 12~query types, broken down by filing jurisdiction. We report individual results for the five jurisdictions with at least 100~queries in the benchmark. The remaining eight jurisdictions: French, Italian, Spanish, Dutch, Portuguese, Norwegian, and Czech, have fewer than 100~queries each and are grouped under ROW (Rest of World).

\begin{figure*}[t]
\centering
\includegraphics[width=\textwidth]{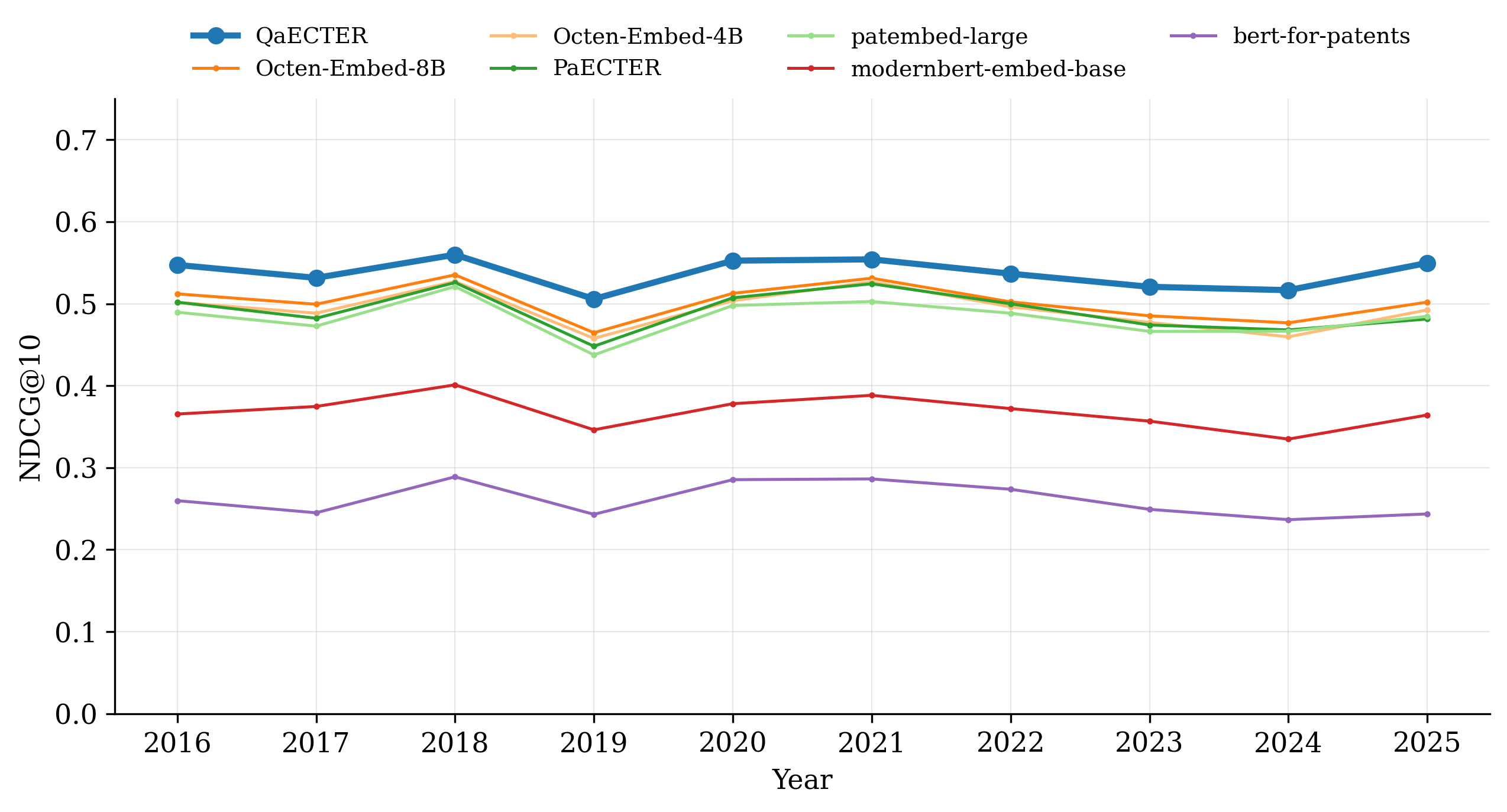}
\caption{NDCG@10 by publication year on \textsc{Sophia-bench} for the TACD query type. All models show stable performance across publication years, with no significant temporal degradation.}
\label{fig:ndcg-by-year}
\end{figure*}

QaECTER achieves the highest NDCG@10 across all six jurisdiction groups. Its advantage holds consistently whether evaluated on Chinese, Japanese, and Korean patents, English-origin filings, or European jurisdictions such as Germany and ROW. Moreover, QaECTER exhibits lower performance variation across jurisdictions than competing models: while others show steeper declines as retrieval difficulty increases, QaECTER maintains a stable margin throughout. This suggests that its cross-jurisdictional robustness is a direct consequence of its training design rather than an artifact of strong results on any single patent corpus.

Performance remains stable across publication years for all models. Notably, QaECTER maintains a consistent lead throughout, and its advantage over competing models widens on the most recent patents (2024--2025). While the full benchmark was excluded from training, the training corpus only covers publications from 2002 to 2024, meaning 2025 patents are entirely unseen, yet our model achieves its strongest relative performance on this subset (Figure~\ref{fig:ndcg-by-year}).

% ──────────────────────────────────────────────────────────
\subsection{Robustness Across Technology Domains}
\label{sec:ipc-results}

\begin{figure*}[t]
\centering
\includegraphics[width=\textwidth]{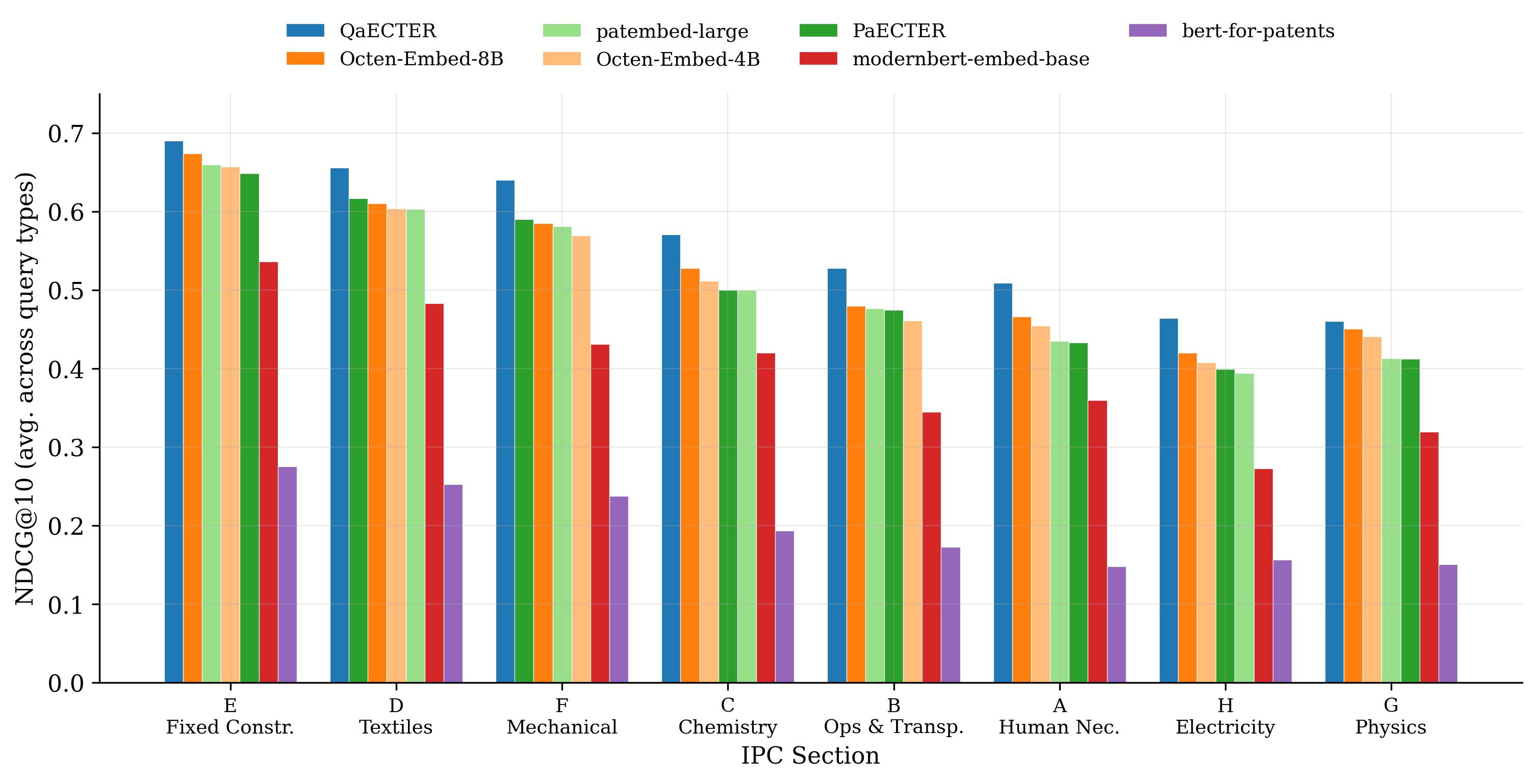}
\caption{NDCG@10 by IPC section on \textsc{Sophia-bench}, averaged across all 12 query types. Sections are sorted by QaECTER performance (descending).}
\label{fig:ndcg-by-ipc}
\end{figure*}

Figure~\ref{fig:ndcg-by-ipc} reports NDCG@10 by IPC section, averaged across query types.

QaECTER ranks first in all eight IPC sections, delivering the best performance across the full range of patent technology domains. Its results remain consistently strong regardless of domain size or technical focus, demonstrating stable and reliable retrieval quality throughout the benchmark. This uniform leadership confirms that QaECTER’s advantage is broad and systematic, rather than concentrated in specific or easier areas, establishing it as a robust general-purpose patent retrieval model.
% ──────────────────────────────────────────────────────────
\begin{table*}[t]
\centering
\caption{InScope analysis on \textsc{Sophia-bench}. Best result per column is in \textbf{bold}.}
\label{tab:inscope}
\small
\setlength{\tabcolsep}{4pt}
\begin{tabular}{ll cc cc cc}
\toprule
& & \multicolumn{2}{c}{\textbf{Section}} & \multicolumn{2}{c}{\textbf{Subclass}} & \multicolumn{2}{c}{\textbf{Subgroup}} \\
\cmidrule(lr){3-4} \cmidrule(lr){5-6} \cmidrule(lr){7-8}
\textbf{Query} & \textbf{Model} & @10 & @100 & @10 & @100 & @10 & @100 \\
\midrule
\texttt{tacd}
& QaECTER (ours)        & \textbf{.915} & .844 & \textbf{.784} & .628 & \textbf{.490} & .297 \\
& Octen-Embed-8B        & .901 & .818 & .762 & .592 & .472 & .279 \\
& Octen-Embed-4B        & .904 & .828 & .765 & .606 & .474 & .287 \\
& PaECTER               & .902 & .826 & .760 & .597 & .465 & .277 \\
& patembed-large        & .914 & \textbf{.851} & .781 & \textbf{.636} & .484 & \textbf{.299} \\
& modernbert-embed-base & .889 & .813 & .730 & .567 & .430 & .250 \\
& bert-for-patents      & .834 & .744 & .614 & .454 & .330 & .187 \\
\midrule
\texttt{ai\_feat.}
& QaECTER (ours)        & \textbf{.913} & .842 & \textbf{.781} & .627 & \textbf{.487} & \textbf{.297} \\
& Octen-Embed-8B        & .901 & .815 & .758 & .590 & .468 & .278 \\
& Octen-Embed-4B        & .901 & .822 & .759 & .599 & .467 & .282 \\
& PaECTER               & .899 & .823 & .751 & .592 & .455 & .274 \\
& patembed-large        & .911 & \textbf{.848} & .776 & \textbf{.633} & .474 & .296 \\
& modernbert-embed-base & .885 & .811 & .720 & .566 & .414 & .246 \\
& bert-for-patents      & .824 & .750 & .599 & .462 & .303 & .187 \\
\bottomrule
\end{tabular}
\end{table*}

\subsection{Domain Relevance (InScope)}
\label{sec:inscope-results}

Table~\ref{tab:inscope} reports InScope at $k \in \{10, 100\}$ across three IPC granularity levels: section (8~broad domains), subclass (4-character codes), and subgroup (full IPC codes).

QaECTER achieves the highest InScope@10 at all three granularity levels, indicating that its top-ranked results are the most topically relevant. At the section level, over 91\% of the top-10 retrieved documents share the query's broad technology domain.

At @100, patembed-large slightly edges ahead (by less than 1 percentage point), but QaECTER leads at @10, indicating stronger precision in the top ranks. This is notable because QaECTER was never trained on classification tasks, yet it matches and surpasses patembed-large, which explicitly included IPC classification during training, highlighting QaECTER's strong generalization beyond its training objectives.

Performance drops substantially from section to subgroup granularity: InScope@10 falls from approximately 0.91 to 0.49. This steep decline reflects the inherent difficulty of fine-grained IPC matching as retrieving documents that share the same specific subgroup code requires capturing highly detailed technical distinctions that are challenging even for specialized models. In practice, combining our model with an IPC-based pre-filter can fully address this limitation.

% ──────────────────────────────────────────────────────────

\begin{table}[t]
\centering
\caption{DAPFAM.ALL benchmark (NDCG@100). Results for models other than QaECTER are from~\cite{patembed}.}
\label{tab:dapfam}
\small
\begin{tabular}{lcc}
\toprule
\textbf{Model} & \textbf{No Prompt} & \textbf{Best} \\
\midrule
QaECTER (ours)       & \textbf{.379} & \textbf{.379} \\
patembed-large       & .044 & .377 \\
PatEmbed-Base        & .352 & .370 \\
PaECTER              & .343 & .343 \\
BGE-PatentMatch      & .314 & .314 \\
bert-for-patents     & .228 & .228 \\
\bottomrule
\end{tabular}
\end{table}
\subsection{External Validation: DAPFAM}
\label{sec:dapfam}

To assess generalization beyond \textsc{Sophia-bench}, we evaluate QaECTER on DAPFAM~\cite{patembed}, an independent patent retrieval benchmark introduced alongside the PatEmbed model family. DAPFAM evaluates cross-view retrieval across six query–corpus configurations, combining title+abstract and title+abstract+claims as queries against three corpus representations. We report the average NDCG@100 across all six DAPFAM.ALL subtasks.

Results are outlined in Table~\ref{tab:dapfam}. QaECTER achieves 0.379 NDCG@100 without any instruction prefix, surpassing every model in the DAPFAM comparison, including patembed-large's best score of 0.377, which requires task-specific prompt prefixes. The result is particularly striking for patembed-large, whose performance collapses to 0.044 without its prompt, revealing a strong dependency on instruction tuning. This strong dependency makes it impossible to use a single corpus embedding across tasks, which is a real burden in large scale production settings. 

In contrast, QaECTER's prompt-free design and strong generalization capabalities makes it directly usable in production systems without per-task prompt engineering. Another important observation here is that on our independent benchmark sophia-bench, PaECTER has nearly identical retrieval results to patembed-large, surpassing it by a thin margin in most tasks, while on DAPFAM patembed-large scores higher than PaECTER, although the tasks in DAPFAM are not structurally different from the tasks of sophia-bench (both tasks are citation-based patent retrieval). This suggests the possibility of overfitting from patembed-large on their DAPFAM test data. 

These results confirm that QaECTER's advantage is not specific to \textsc{Sophia-bench} but extends to independently constructed benchmarks with different ground truth and evaluation protocols establishing itself as the new state of the art patent embeddings model.

\section{Conclusion}

We presented two complementary contributions to patent retrieval. \textsc{Sophia-bench} provides the first large-scale benchmark that systematically evaluates embedding models across 12~query representations, eight technology domains, multiple jurisdictions, and a decade of patent filings. By combining citation-based ground truth with the InScope domain-relevance metric, it enables fine-grained diagnosis of model strengths and failure modes that prior benchmarks could not capture.

\textsc{QaECTER}, our 344M-parameter embedding model, establishes a new state of the art on both \textsc{Sophia-bench} and the independent DAPFAM benchmark. Its consistent superiority over models up to 23$\times$ larger demonstrates that domain-specific training on patent citation graphs with multi-view self-alignment can be far more effective than scaling general-purpose architectures. Critically, QaECTER achieves this without instruction prompts or task-specific prefixes, making it directly deployable in production search pipelines.

Our evaluation also yields several broader findings. AI-generated summaries match or exceed traditional patent fields as queries, validating their use in modern search workflows. The InScope analysis reveals that even the best models struggle with fine-grained IPC subgroup matching, pointing to an important direction for future work.

\newpage
\bibliographystyle{IEEEtran}
\bibliography{refs}

\end{document}